# Novel LDPC Decoder via MLP Neural Networks

A. R. Karami[1], Electrical Engineering Department, Dalhousie University

M. Ahmadian Attari[2], Faculty of Electrical Engineering, K. N. Toosi University of Technology

Email: AKarami@dal.ca[1]    m_ahmadian@kntu.ac.ir[2]

**Abstract**: In this paper, a new method for decoding Low Density Parity Check (LDPC) codes, based on Multi-Layer Perceptron (MLP) neural networks is proposed. Due to the fact that in neural networks all procedures are processed in parallel, this method can be considered as a viable alternative to Message Passing Algorithm (MPA), with high computational complexity. Our proposed algorithm runs with soft criterion and concurrently does not use probabilistic quantities to decide what the estimated codeword is. Although the neural decoder performance is close to the error performance of Sum Product Algorithm (SPA), it is comparatively less complex. Therefore, the proposed decoder emerges as a new infrastructure for decoding LDPC codes.

**Key Words**: LDPC codes, Tanner graph, Iterative decoding, MPA, MLP neural networks, Gradient descent algorithm

## I- Introduction

LDPC codes were introduced by Gallager in 1962 [1]. These codes are a type of block codes, being so considerable because of their near to ideal performance known as Shannon limit. In addition to the construction of LDPC codes, the paper presented an iterative algorithm for decoding LDPC codes. However, the complexity of the algorithm was higher than the power of existent electronic processors. For this reason, despite Tanner's effort to revive LDPC codes in 1981 [2], these codes have been forgotten until 1996, when MacKay and Neal rediscovered LDPC codes as a competitor to turbo codes [3, 4]. A LDPC code is represented with its sparse parity check matrix and the corresponding Tanner graph.

LDPC codes can be considered serious competitors to turbo codes in terms of performance and complexity [5]. However, much of the work on LDPC decoder design has been directed towards achieving optimal tradeoffs between complexity and coding gain [6, 7].

Using neural networks as an alternative for decoding block and convolutional codes has already been introduced in some papers. In [8] a neural network for Additive White Gausian Noise (AWGN) channel is proposed to decode a Hamming (7, 4) code. The decoder is designed so that it guarantees the Maximum Likelihood Decoding (MLD). In [9] a neural decoder for block codes which has a better performance than Hard Decision Decoding (HDD) is introduced. The decoder in [10] includes a neural network which has been trained by syndromes and its output is actually an error vector that must be added to the received sequence to correct it. In [11] the decoder has N-1 (N is the number of code words) neurons in output layer and n (code length) neurons for input layer. The network must be trained for all code words except for all-zero code word. A neural network decoder of convolutional codes is proposed in [12] as well as [13,14] which estimate bit values based on minimizing noise energy function.

Decoding LDPC codes comprises an iterative approach, in which optimized values of output are to be obtained. We have encountered the problems of optimization and iterative algorithms. In this regard, neural networks are suggested as an important and referable tool to address these problems. Their optimization capabilities and their iterative structure are capable for developing a new method for decoding LDPC codes. The proposed algorithm not only makes a new foundation of decoding LDPC codes, but also can provide low complex block and convolutional decoders. The proposed algorithm is compared to original SPA not to its simplified version, min-sum algorithm.

This paper is organized as follows. In Section II, decoding methods of LDPC codes have been briefly reviewed. In Section III, a short explanation of MLP neural networks has been provided, and then our proposed structure and method of neural decoding have been elaborated. Section IV devoted to complexity calculation of SPA and MLPD. Computer simulation results for MLPD have been included in Sections V. Finally, Section VI concludes the paper.

## II- Decoding Methods for LDPC Codes

LDPC codes can be decoded by means of a variety of decoding algorithms such as Bit Flipping (BF) [1] and its developed variants [15, 16] and MPA that includes SPA and its variants. MPA is indeed near optimal decoding method for LDPC codes [17].

Message passing algorithm includes several algorithms, which use an iterative process of decoding. In any iteration, the messages are transferred between two parts of the corresponding Tanner graph. However, it is necessary to calculate the a posteriori probability (APP) $P_i$.

$$P_i = \Pr(c_i = 1 | r) \tag{1}$$

where $c_i$ is $i$-th bit of the transmitted code word $\mathbf{c} = [c_1, c_2, ..., c_n]$, and word $\mathbf{r} = [r_1, r_2, ..., r_n]$ is the received vector from the channel.

SPA is a soft decoding method for LDPC codes. First, we present the following notations:

- $X_j$ : variable nodes connected to the check node $s_j$
- $Z_i$ : check nodes connected to the variable node $c_i$
- $X_j \setminus i$ : variable nodes connected to the check node $s_j$ excluding $c_i$
- $Z_i \setminus j$ : check nodes connected to the variable node $c_i$ excluding $s_j$
- $M_X(\sim i)$ : messages from all variable nodes excluding $c_i$
- $M_Z(\sim j)$ : messages from all check nodes excluding $s_j$
- $E_i$ : event that parity equations including $c_i$ are satisfied
- $q_{ij}(b)$ : messages from the variable node $c_i$ to check node $s_j$.

$$q_{ij}(b) = \Pr(c_i = b | E_i, r_i, M_Z(\sim j)) \qquad (2)$$
$$b \in \{0,1\}$$

- $r_{ji}(b)$ : messages from check node $s_j$ to variable node $c_i$

$$r_{ji}(b) = \Pr(\text{parity check equation } s_j \text{ is satisfied} | c_i = b, M_X(\sim i)) \qquad (3)$$
$$b \in \{0,1\}$$

The concepts of $q_{ij}(b)$ and $r_{ji}(b)$ are illustrated in Fig. 1.

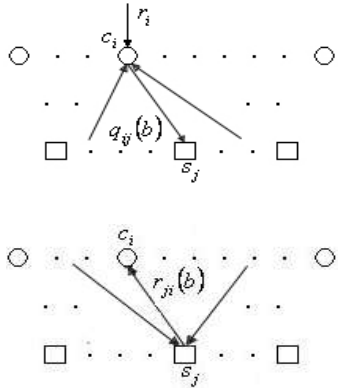

Fig. 1 Passing messages calculate $r_{ji}(b)$ and $q_{ij}(b)$

It can be easily expressed that

$$q_{ij}(0) = (1-P_i)\Pr(E_i|c_i = 0, r_i, M_Z(\sim j))/\Pr(E_i)$$
$$= \gamma_{ij}(1-P_i)\prod_{j' \in Z_i \setminus j} r_{j'i}(0) \tag{4}$$
$$q_{ij}(1) = \gamma_{ij} P_i \prod_{j' \in Z_i \setminus j} r_{j'i}(1)$$

and

$$r_{ji}(0) = \frac{1}{2} + \frac{1}{2}\prod_{i' \in X_j \setminus i}^{M}(1 - 2q_{i'j}(1)) \tag{5}$$
$$r_{ji}(1) = 1 - r_{ji}(0)$$

where $\gamma_{ij}$ is chosen to guarantee $q_{ij}(1) + q_{ij}(0) = 1$. Hence, by any iteration $\gamma_{ij}$ is calculated using (6).

$$\gamma_{ij} = \frac{1}{(1-P_i)\prod_{j' \in Z_i \setminus j} r_{j'i}(0) + P_i \prod_{j' \in Z_i \setminus j} r_{j'i}(1)} \tag{6}$$

The ordinary SPA is presented as follows.

a. For $i = 1,...,n$ determine $P_i = \Pr(c_i = 1|r_i)$. Then determine $q_{ij}(0) = 1 - P_i$ and $q_{ij}(1) = P_i$ for any $i, j$ that $h_{ij} = 1$.

b. Bring up to date $r_{ji}(0)$, and $r_{ji}(1)$ using (5).

c. Bring up to date $q_{ij}(0)$, and $q_{ij}(1)$ using (4).

d. For $i = 1,...,n$ calculate

$$Q_i(0) = K_i(1-P_i)\prod_{j \in Z_i} r_{ji}(0)$$
$$Q_i(1) = K_i P_i \prod_{j \in Z_i} r_{ji}(1) \tag{7}$$

The constants $K_i$ are chosen to guarantee $Q_i(1) + Q_i(0) = 1$.

e. For $i = 1,...,n$ set

$$\hat{c}_i = \begin{cases} 1 & Q_i(1) > Q_i(0) \\ 0 & else \end{cases} \quad (8)$$

If equation $\hat{c}.\mathbf{H}^T = 0$ is satisfied or number of iterations meets its maximum limit, stop; otherwise, go to step b.

## III- The Proposed Neural Decoder for LDPC Codes

As indicated in Fig. 2, a neural network is created by placing the neurons in different layers and then connecting the outputs of the neurons of a layer to the inputs of the neurons in the next layer.

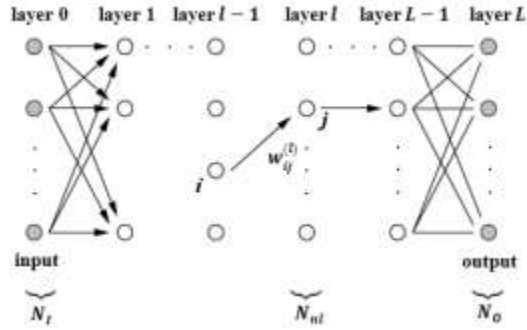

Fig. 2   Structure of a MLP neural network

The output of layer $l$ of the MLP neural network is obtained from (9).

$$O^{(l)} = f^{(l)}\left(W^{(l)^T} O^{(l-1)} + B^{(l)}\right) \quad (9)$$

where $W^{(l)}$ is the weight matrix of layer $l$, $B^{(l)}$ is the bias matrix of layer $l$, $O^{(l-1)}$ is the output matrix of layer $l-1$ or the input matrix of layer $l$ and $W^{(l)^T}$ is the transpose of $W^{(l)}$.

In a neural network, weights and biases are adjustable parameters. They can be adjusted based on a set of given data. The process of finding and adjusting the weights and biases of a neural network is called training. The purpose of neural network training is to reduce the Sum Square Error (SSE) function $E$, defined as:

$$E = \frac{1}{2}\sum_{j=1}^{N_O}(e_j)^2 \tag{10}$$

where $e_j$ is the error of $j$-th entry of the output matrix and equals to the difference between the desired value and actual value of the output. Required modifications in the training parameters may be considered using optimization algorithms such as gradient descent algorithm [18].

$$\Delta W^{(l)}(n) = -\mu_{W^{(l)}}\frac{\partial E(n)}{\partial W^{(l)}(n)} \qquad W^{(l)}(n) = W^{(l)}(n-1) + \Delta W^{(l)}(n)$$

$$\Delta B^{(l)}(n) = -\mu_{B^{(l)}}\frac{\partial E(n)}{\partial B^{(l)}(n)} \qquad B^{(l)}(n) = B^{(l)}(n-1) + \Delta B^{(l)}(n)$$

where both $\mu_{W^{(l)}}$ and $\mu_{B^{(l)}}$ are constant values between 0 and 1, and are called training rate of $W^{(l)}$ and training rate of $B^{(l)}$, respectively.

Our presented decoder [19], which is a MLP neural network has 2 layers and has been formed based on the Tanner graph of the LDPC codes. The structure of the presented neural network for decoding LDPC codes with any length and characters is shown in Fig. 3. As indicated in the figure, this network is similar to the corresponding Tanner graph.

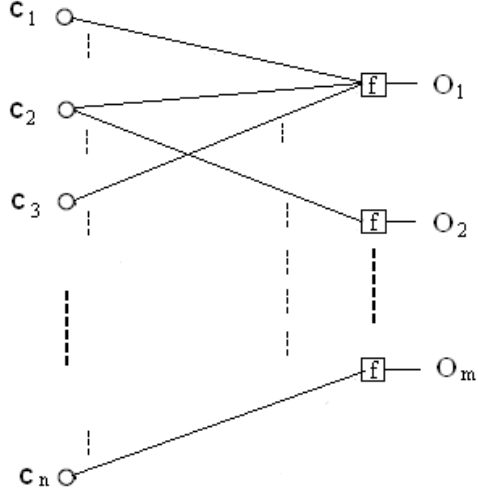

**Fig. 3 The structure of the proposed neural decoder**

In this model, layer 1 corresponds to the variable nodes and layer 2 corresponds to the check nodes of Tanner graph. Therefore, if the input of layer 1 is a code word, then the desired values in the output of layer 2 are zeros. The function of the output layer of the network must be the XOR function in analog conditions. This function with two variables is given in (11).

$$x \oplus y = x(1-y) + y(1-x) \tag{11}$$

However, in practice, the number of code word components, corresponding to the number of received vector components is much more than two. Therefore, XOR function of the related variables can be expressed in the form of (12).

$$\begin{aligned} x_1 \oplus x_2 \oplus ... \oplus x_m &= (((x_1 \oplus x_2) \oplus x_3) \oplus ...) \oplus x_m \\ &= ((x_1' \oplus x_3) \oplus ...) \oplus x_m \\ &= (x_1'' \oplus ...) \oplus x_m \end{aligned} \tag{12}$$

where
$$x_1 \oplus x_2 = x_1'$$
$$x_1' \oplus x_3 = x_1''$$

The explained XOR function is differentiable, and so it can be used in the training process of the neural network using gradient descent algorithm.

The components of the output vector $\mathbf{O} = [o_1, o_2, ... o_m]$ correspond to the check nodes of Tanner graph. Accordingly, the network parameters are trained so that $\mathbf{O}$ tends to zero. The SSE function $E$ of the network is calculated in (13).

$$E = \frac{1}{2}\sum_{j=1}^{m}(e_j)^2 = \frac{1}{2}\sum_{j=1}^{m}(0-o_j)^2 = \frac{1}{2}\sum_{j=1}^{m}(o_j)^2 \qquad (13)$$

The training parameters of our proposed decoder are the received vector components or the inputs of the network. The components change as far as the SSE function $E$ approaches its minimum point. Finally, after training the input vector, the components of the altered vector are mapped to 0 or 1 depending on the minimum Euclidean distance to 0 and 1. The created word is an estimation of the transmitted code word. In any iteration, the necessary alteration of the inputs is expressed in (14).

$$\Delta c_i = -\mu \frac{\partial E}{\partial c_i} = -\mu \sum_{j=1}^{m}\left(\frac{\partial E}{\partial e_j}\frac{\partial e_j}{\partial o_j}\frac{\partial o_j}{\partial c_i}\right) \qquad (14)$$

where $\mu$ is the training rate of the input vector. Choosing proper value for $\mu$ is an important factor which affects the performance of the neural decoder. This selection should be in a way that the probability of placing $E$ in the domain of local minimum points decreases. From (13), we have:

$$\frac{\partial E}{\partial e_j} = e_j$$
$$\frac{\partial e_j}{\partial o_j} = -1 \qquad (15)$$

and then

$$\Delta c_i = \mu \sum_{j=1}^{m} \left( e_j \frac{\partial o_j}{\partial c_i} \right) = -\mu \sum_{j=1}^{m} \left( o_j \frac{\partial o_j}{\partial c_i} \right) \tag{16}$$

$\frac{\partial o_j}{\partial c_i}$ is a function of the values of those input nodes ( variable nodes ) that along with $c_i$ create the XOR function of the output node ( check node ) $o_j$. However, considering $I_j \setminus i$ as the input nodes connected to the output node $o_j$ except $c_i$, we will prove (17) as a Lemma.

$$\frac{\partial o_j}{\partial c_i} = \prod_{l \in I_j \setminus i}(1-2c_l) \tag{17}$$

**Proof:** Assume in a typical output node $o_t$, input nodes $a_1, a_2,..., a_m$ cooperate. Then,

$$o_t = a_1 \oplus a_2 \oplus ... \oplus a_m \tag{18}$$

We define

$$\begin{aligned}
a_1 &= b_1 \\
a_1 \oplus a_2 &= b_2 \\
a_1 \oplus a_2 \oplus a_3 &= b_2 \oplus a_3 = b_3 \\
&... \\
a_1 \oplus a_2 \oplus a_3 \oplus ... \oplus a_z &= b_{z-1} \oplus a_z = b_z
\end{aligned} \tag{19}$$

Regarding the XOR function definition, $b_i$ s can be calculated from (20).

$$\begin{aligned}
a_1(1-a_2) + a_2(1-a_1) &= b_2 \\
b_2(1-a_3) &= a_3(1-b_2) = b_3 \\
&... \\
b_{z-1}(1-a_z) + a_z(1-b_{z-1}) &= b_z
\end{aligned} \tag{20}$$

For the input node $a_1$ we have:

$$\frac{\partial o_t}{\partial a_1} = \frac{\partial o_t}{\partial b_{m-1}} \frac{\partial b_{m-1}}{\partial b_{m-2}} \cdots \frac{\partial b_3}{\partial b_2} \frac{\partial b_2}{\partial a_1} \tag{21}$$

With reference to (20), it is obvious that

$$\frac{\partial b_z}{\partial b_{z-1}} = 1 - 2a_z \tag{22}$$

Considering the above definitions, it can be stated that $o_t = b_m$. Therefore, from (22) we can rewrite (21) in the form of (23).

$$\begin{aligned}\frac{\partial o_t}{\partial a_1} &= (1-2a_m)(1-2a_{m-1})\ldots(1-2a_3)(1-2a_2) \\ &= \prod_{\substack{l=1 \\ l \neq 1}}^{m}(1-2a_l)\end{aligned} \tag{23}$$

It is clear that there is no difference between place of $a_1$ and places of other $a_i$s in XOR function. Thus,

$$\begin{aligned}\frac{\partial o_t}{\partial a_2} &= (1-2a_m)(1-2a_{m-1})\ldots(1-2a_3)(1-2a_1) \\ &= \prod_{\substack{l=1 \\ l \neq 2}}^{m}(1-2a_l)\end{aligned} \tag{24}$$

...

and in a general form:

$$\begin{aligned}\frac{\partial o_t}{\partial a_i} &= (1-2a_m)(1-2a_{m-1})\ldots(1-2a_2)(1-2a_1) \\ &= \prod_{\substack{l=1 \\ l \neq i}}^{m}(1-2a_l)\end{aligned} \tag{25}$$

■

## IV- Complexity of the Decoders

In this section, the proposed algorithm and SPA are fully investigated and their complexities are calculated.

Since arithmetic order of multiplication is higher than that of summation, nothing is lost if only the number of multiplications in each algorithm is calculated and compared.

## IV-1- Complexity of SPA

According to Section II, any iteration of SPA involves two half-iterations. Assume $N^{\Pi}_{q_{ij}(0)}$ and $N^{\Pi}_{q_{ij}(1)}$ to be the number of multiplications required for the computation of $q_{ij}(0)$ and $q_{ij}(1)$, respectively. Therefore, regarding (4) and the fact that $N^{\Pi}_{q_{ij}(0)} = N^{\Pi}_{q_{ij}(1)}$, in the first half-iteration the number of multiplications $N^{\Pi}_{q}$ is attained by (26).

$$N^{\Pi}_{q} = 2\sum_{i=1}^{n} N^{\Pi}_{q_i(0)} \qquad (26)$$

where $N^{\Pi}_{q_i(0)}$ is the half number of multiplications related to the variable node $c_i$ and is obtained using (27).

$$N^{\Pi}_{q_i(0)} = \sum_{j=1}^{n-k} N^{\Pi}_{q_{ij}(0)} = \sigma_i^2 \qquad (27)$$

where $\sigma_i$ is number of 1s in $i$–th column of matrix **H**. For a regular code with $\sigma$ as number of 1s in each column of matrix **H**, $N^{\Pi}_{q} = 2n\sigma^2$.

For the second half-iteration suppose $N^{\Pi}_{r_{ji}(0)}$ and $N^{\Pi}_{r_{ji}(1)}$ are the number of multiplications required to compute $r_{ji}(0)$ and $r_{ji}(1)$, respectively. Hence, considering (5) and the fact that $r_{ji}(1) = 1 - r_{ji}(0)$, in the second half-iteration the number of multiplications $N^{\Pi}_r$ is attained by (28).

$$N^{\Pi}_r = \sum_{j=1}^{n-k} N^{\Pi}_{r_j(0)} + \sum_{j=1}^{n-k} \rho_j \qquad (28)$$

where $N^{\Pi}_{r_j(0)}$ is the total number of multiplications related to variable node $s_j$ and is obtained using (29). The second term of the summation in (28) is attributed to $2q_{i'j}$ in (5) and $\rho_j$ is the number of 1s in $j-$th row of matrix **H.**

$$N^{\Pi}_{r_j(0)} = \sum_{i=1}^{n} N^{\Pi}_{rji(0)} = (\rho_j - 1)\rho_j \qquad (29)$$

Eventually, by inserting (29) in (28), (30) represents $N^{\Pi}_r$.

$$N^{\Pi}_r = \sum_{j=1}^{n-k} \rho_j^2 \qquad (30)$$

For a regular code with $\rho$ as the number of 1s in each row of matrix **H,** $N^{\Pi}_q = (n-k)\rho^2$. According to SPA, for decision if related bit is 1 or 0, there are other multiplications, which are disclosed in (7); we exhibit their number by $N^{\Pi}_{Q(0)}$ and $N^{\Pi}_{Q(1)}$. Obviously, $N^{\Pi}_{Q(0)} = N^{\Pi}_{Q(1)}$. Therefore, the total number of multiplications, related to clause *d* of SPA $N^{\Pi}_Q$ calculated as follows.

$$N^{\Pi}_Q = 2\sum_{i=1}^{n} (\sigma_i + 1) \qquad (31)$$

For a regular code with $\sigma$ as the number of 1s in each column of matrix **H**, $N_Q^\Pi = 2n(\sigma+1)$. Finally, total number of multiplications in any iteration of SPA $N_{SPA}^\Pi$ can be clarified in (32).

$$N_{SPA}^\Pi = 2\sum_{i=1}^{n}\left(\sigma_i^2 + \sigma_i + 1\right) + \sum_{j=1}^{n-k}\rho_j^2 \tag{32}$$

## IV-2- Complexity of the Proposed Decoder

With reference to Section III and alike to SPA, MLPD has two half-iterations. In the first half-iteration some XOR functions are executed. We try to form some relation to calculate the number of multiplications, required for these XOR functions. In this regard, assuming $N_\oplus^\Pi$ it can be written as:

$$N_\oplus^\Pi = \sum_{j=1}^{n-k} N_{\oplus f_j}^\Pi \tag{33}$$

$N_{\oplus f_j}^\Pi$ is number of multiplications, required to do XOR function in $j$th output neuron. It is calculated in (34) by using (11) and (12).

$$N_{\oplus f_j}^\Pi = 2(\rho_j - 1) \tag{34}$$

For a regular code with $\rho$ the number of 1s in each row of matrix **H**, $N_{\oplus f_j}^\Pi = 2(n-k)(\rho - 1)$.

For the second half-iteration (16) is used. First, the total number of multiplications, which are related to $\dfrac{\partial o_j}{\partial c_i}$, and are done in $f_j$ must be described named as $N_{\partial o_j}^\Pi$.

$$N_{\partial o_j}^{\Pi} = (\rho_j - 2)\rho_j \tag{35}$$

The total number of multiplications to attain needed changes for network input $\mathbf{c} = [c_1, c_2, ..., c_n]$ is $N_{\Delta c}^{\Pi}$ which is explained in (36).

$$N_{\Delta c}^{\Pi} = \sum_{j=1}^{n-k} \left( N_{\partial o_j}^{\Pi} + 2\rho_j \right) + n \tag{36}$$

where $n$ is related to $1 - 2c_{i'}$ which appears in (17). Equation (36) can be rewritten as follows.

$$N_{\Delta c}^{\Pi} = \sum_{j=1}^{n-k} \left( \rho_j^2 \right) + n \tag{37}$$

For a regular code with $\rho$ as the number of 1s in each row of matrix **H**, $N_{\Delta c}^{\Pi} = (n-k)\rho^2 + n$. Finally, the number of multiplications in any iteration of MLPD $N_{MLPD}^{\Pi}$ can be presented by (38).

$$N_{MLPD}^{\Pi} = \sum_{j=1}^{n-k} \left( \rho_j^2 + 2\rho_j - 2 \right) + n \tag{38}$$

### IV-3- Complexity Comparison of the Decoders

Regarding the fact that $\sum_{i=1}^{n} \sigma_i = \sum_{j=1}^{n-k} \rho_j$, (32) can be rewritten as follows.

$$N_{SPA}^{\Pi} = \sum_{j=1}^{n-k} \left( \rho_j^2 + 2\rho_j \right) + 2\sum_{i=1}^{n} \sigma_i^2 + 2n = N_{MLPD}^{\Pi} + 2\sum_{i=1}^{n} \sigma_i^2 + 3n - 2k \tag{39}$$

Equation (39) proves how much the complexity of MLPD is smaller than that of SPA. However, increasing code length and also density of parity check matrix affect on robustness of the proposed decoder. The former has a linear effect on difference of complexity between the two decoders. The latter is in square relationship with that.

## V- Simulation Results and Performance of the proposed Decoder

For simulation of the proposed algorithm, we consider two examples, embracing two different codes.

**Example 1:** The considered code is a (20, 1, 2) LDPC code.

At the first step, the received vector from the channel enters the input layer of the network. Then, through any iteration in the output layer of the network, the XOR function of the related nodes of the input layer is calculated. After that, the SSE function $E$ is calculated based on (13). Using (14) the necessary changes of the input layer components are performed and then added to the previous values of these components. In probable next iterations, this approach would continue until the SSE function $E$ significantly decreases, or the number of iterations exceeds a certain threshold.

Fig. 4 compares the performances of MLPD and SPA and shows that by keeping the same error rate it is possible to utilize a simpler and faster decoder.

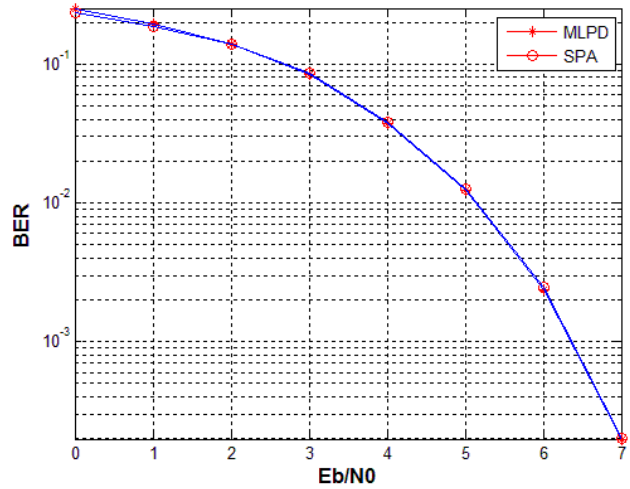

**Figure 4 The BER diagrams of MLPD and SPA**

Complexity calculations of SPA and MLPD for the code based on (32) and (38) are 160 and 80, respectively which shows that MLPD is less complex compared with SPA.

**Example 2:** The considered code is LDPC(60, 1, 3) code.

The same procedure as in the prior example takes place. Finally, Fig. 5 shows how much close are the two decoders' performances. As in the previous example, the MLPD is here a good competitor to SPA while it is simpler and faster with a tight performance result.

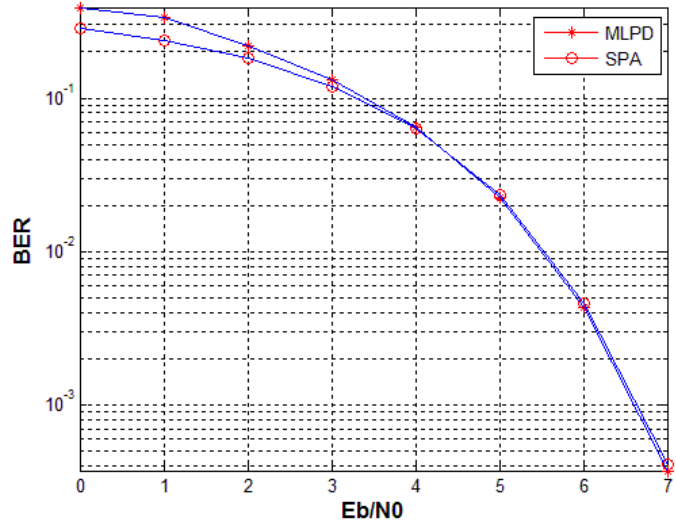

Fig. 5 The BER diagrams of MLPD and SPA

Using (32) and (38) the complexity criterion of SPA and MLPD for the code are 540 and 320, respectively which performs less complexity of MLPD compared to SPA for this code.

## VI- Conclusion

In this article, a new decoder of LDPC codes has been proposed based on the neural networks. Also, we have developed a method to calculate the complexity of SPA and the proposed algorithm. The proposed decoder is a soft decision decoder for LDPC codes. The neural decoder is based on the Tanner graph and can be considered as a type of massage passing algorithm, where the transferred massages are not probabilistic amounts. In this context, the need for comparing the calculated probabilities for each situation and using memory for these probabilities can be ignored. The proposed decoder operates with less complexity than of SPA. Moreover, comparison of the proposed algorithm with SPA shows similar performance results. Selection of suitable training rates and other functions in the output layer of the neural decoder, improvement in the training of the neural decoder, and application of other optimization methods can be considered to improve the performance and speed of the proposed decoder.

## Acknowledgement

This research is partly supported by Iran National Science Foundation (INSF).